\documentclass[fleqn,usenatbib]{mnras}

\usepackage{newtxtext,newtxmath}

\usepackage[T1]{fontenc}
\usepackage{ae,aecompl}

\usepackage{graphicx}	
\usepackage{amsmath}	
\usepackage{amssymb}	
\usepackage{mathtools}

\title[Planets from ring-like structures in discs]{The newborn planet population emerging from ring-like structures in discs}

\author[Lodato et al.]{\parbox{\textwidth}{
Giuseppe Lodato$^{1}$\thanks{E-mail: giuseppe.lodato@unimi.it},
Giovanni Dipierro$^{2}$,
Enrico Ragusa$^{1,2}$,
Feng Long$^{3}$, Gregory J. Herczeg$^{3}$,
Ilaria Pascucci$^{4,5}$,
Paola Pinilla$^{6}$,
Carlo F. Manara$^{7}$,
Marco Tazzari$^{8}$,
Yao Liu$^{9,10}$,
Gijs D. Mulders$^{5,11}$,
Daniel Harsono$^{12}$,
Yann Boehler$^{13}$,
Fran\c cois M\'enard$^{13}$,
Doug Johnstone$^{14,15}$,
Colette Salyk$^{16}$,
Gerrit van der Plas$^{13}$,
Sylvie Cabrit$^{17}$,
Suzan Edwards$^{18}$,
William J. Fischer$^{9}$,
Nathan Hendler$^{4}$,
Brunella Nisini$^{20}$,
Elisabetta Rigliaco$^{21}$,
Henning Avenhaus$^{9}$,
Andrea Banzatti$^{4}$,
Michael Gully-Santiago$^{22}$
}\vspace{0.15cm}\\
$^{1}$Dipartimento di Fisica, Universit\`a degli Studi di Milano, Via Celoria 16, 20133 Milano, Italy\\
$^{2}$Department of Physics and Astronomy, University of Leicester, Leicester, LE1 7RH, United Kingdom\\
$^{3}$Kavli Institute for Astronomy and Astrophysics, Peking University, Yiheyuan 5, Haidian Qu, 100871 Beijing, China\\
$^{4}$ Lunar and Planetary Laboratory, University of Arizona, Tucson, AZ 85721, USA\\
$^5$ Earths in Other Solar Systems Team, NASA Nexus for Exoplanet System Science, USA\\
$^6$ Department of Astronomy/Steward Observatory, The University of Arizona, 933 North Cherry Avenue, Tucson, AZ 85721, USA\\
$^7$European Southern Observatory, Karl-Schwarzschild-Str. 2, D-85748 Garching bei M\"{u}nchen, Germany\\
$^8$ Institute of Astronomy, University of Cambridge, Madingley Road, Cambridge CB3 0HA, UK\\
$^9$ Max Planck Institute for Astronomy, K\"onigstuhl 17, D-69117 Heidelberg, Germany\\
$^{10}$Purple Mountain Observatory \& Key Laboratory for Radio Astronomy, Chinese Academy of Sciences,\\
2 West Beijing Road, Nanjing 210008, China\\
$^{11}$
Department of the Geophysical Sciences, The University of Chicago, Chicago, IL 60637, USA\\
$^{12}$Leiden Observatory, Leiden University, P.O. box 9513, 2300 RA Leiden, The Netherlands\\
$^{13}$Univ. Grenoble Alpes, CNRS, IPAG, F-38000 Grenoble, France\\
$^{14}$ NRC Herzberg Astronomy and Astrophysics, 5071 West Saanich Road, Victoria, BC, V9E 2E7, Canada\\
$^{15}$Department of Physics and Astronomy, University of Victoria, Victoria, BC, V8P 5C2, Canada\\
$^{16}$Vassar College Physics and Astronomy Department, 124 Raymond Avenue, Poughkeepsie, NY 12604, USA\\
$^{17}$Sorbonne Universit\'{e}, Observatoire de Paris, Universit\'{e} PSL, CNRS, LERMA, F-75014 Paris, France\\
$^{18}$Five College Astronomy Department, Smith College, Northampton, MA 01063, USA\\
$^{19}$Space Telescope Science Institute Baltimore, MD 21218, USA\\
$^{20}$INAF-Osservatorio Astronomico di Roma, via di Frascati 33, 00040 Monte Porzio Catone, Italy\\
$^{21}$INAF-Osservatorio Astronomico di Padova, Vicolo dell'Osservatorio 5, 35122 Padova, Italy \\
$^{22}$NASA Ames Research Center and Bay Area Environmental Research Institute, Moffett Field, CA 94035, USA\\
}
\date{Accepted XXX. Received YYY; in original form ZZZ}

\pubyear{2019}

\begin{document}
\label{firstpage}
\pagerange{\pageref{firstpage}--\pageref{lastpage}}
\maketitle

\begin{abstract}
ALMA has observed a plethora of ring-like structures in planet forming discs at distances of 10-100 au from their host star. Although several mechanisms have been invoked to explain the origin of such rings, a common explanation is that they trace new-born planets. Under the planetary hypothesis, a natural question is how to reconcile the apparently high frequency of gap-carving planets at 10-100 au with the paucity of Jupiter mass planets observed around main sequence stars at those separations. Here, we provide an analysis of the new-born planet population emerging from observations of gaps in discs, under the assumption that the observed gaps are due to planets. We use a simple estimate of the planet mass based on the gap morphology, and apply it to a sample of gaps recently obtained by us in a survey of Taurus with ALMA. We also include additional data from recent published surveys, thus analysing the largest gap sample to date, for a total of 48 gaps. The properties of the purported planets occupy a distinctively different region of parameter space with respect to the known exo-planet population, currently not accessible through planet finding methods. Thus, no discrepancy in the mass and radius distribution of the two populations can be claimed at this stage. We show that the mass of the inferred planets conforms to the theoretically expected trend for the minimum planet mass needed to carve a dust gap. Finally, we estimate the separation and mass of the putative planets after accounting for migration and accretion, for a range of evolutionary times, finding a good match with the distribution of cold Jupiters.
\end{abstract}

\begin{keywords}
accretion, accretion discs -- planets and satellites: formation -- protoplanetary discs
\end{keywords}



\section{Introduction}

The discovery of the HL Tau disc and its system of rings \citep{Alma15} has marked a new era in our understanding of the gas and dust discs around young stellar objects. Disc substructures appear to be commonplace, and in particular, the most frequently observed structures are  regular, almost axisymmetric rings  \citep{Andrewsetal16,Isellaetal16,Fedeleetal17,Fedeleetal18,Dipierroetal18,Hendleretal18,Clarkeetal18,Vanter18,Liuprep18}. Many theoretical models have been proposed to explain the origin of such rings, including dead zones \citep{Rugeetal16}, condensation fronts \citep{zbb15}, self-induced dust pile-ups \citep{Gonzalezetal15}, self-induced reconnection in magnetized disc-wind systems \citep{Suriano18} or large scale vortices \citep{Bargeetal17}. However, another natural explanation is to associate the gap in the disc to the presence of an embedded planet \citep{Longetal18,Huang18}. This hypothesis has been tested extensively by comparing the disc emission obtained from ALMA observations to that computed from detailed hydrodynamical and radiative transfer simulations \citep[e.g.][]{Dipierroetal15,Clarkeetal18}. 

Several questions arise, however, if one assumes a planetary origin for gaps in discs. In particular, gaps are typically observed at radial distances from the star of the order of $10-100$ au \citep{zhang16a}. It is therefore natural to ask how to reconcile this evidence with the lack of Jupiter-mass planets at such distances around main-sequence stars, as apparent from the extensive planet-detection campaigns of the last decade \citep{Bowler18}. In order to understand the orbital and physical evolution of planets from birth to adulthood, we need to compare the properties of planets around T Tauri stars and young stellar objects to those of planets around main-sequence stars. Such a comparison is not easy because usually the planet properties in gapped discs are obtained through complex and time-consuming numerical simulations, which are not feasible for large samples, and are sensitive to several physical parameters (dust-gas coupling, disc thermodynamics, etc.), for which specific assumptions need to be made. 

In this paper, we provide an analysis of the properties of the new-born planet population, as implied from a sample of gaps and rings detected in our recent survey of discs in the Taurus-Auriga star forming region. To this end, we use a simple prescription to relate the observed width of the gap to the mass of planet assumed to be responsible for its opening. We then relate the resulting planetary properties to the stellar properties and to the population of known exo-planets. 


This paper is organised as follows. In Section 2 we describe the simple method we use to give an estimate of the planet mass based on the gap morphologies. In Section 3 we show our main results. In Section 4 we draw our conclusions.

\section{Planet properties from disc gaps}

Recently, \citet{Longetal18} investigated a sub-sample of 12 discs showing substructures within a larger sample of 32 discs in Taurus obtained with ALMA Band 6 (at 1.3 mm) in Cycle 4 (ID: 2016.1.01164.S; PI: Herczeg).
The sample selection will be fully described by Long et al. (in preparation). Briefly, the sample was selected from stars in Taurus with spectral types earlier than M3 and with line-of-sight extinctions $<3$ mag. The selection was unbiased to the disc mm flux and to any previously known disk structures from mid-IR photometry; the primary bias is the exclusion of disks that had been previously imaged with ALMA at high spatial resolution.
Some of these discs show multiple rings and gaps, providing us with a total of 15 gaps with known morphologies (excluding four additional discs with inner cavities). In Table~\ref{tab:gaps} we provide a summary of the gap properties relevant to the present study. A more detailed analysis can be found in \citet{Longetal18}. 

\begin{table*}
	\centering
	\caption{Gap properties used in this study (from \citealt{Longetal18}). The columns indicate, respectively: (1) star name; (2) Gap width over gap location; (3) gap location with uncertainties from \citet{Longetal18}; (4) Stellar mass; (5) total mm-flux at 1.3 mm of source; (6) Total dust mass from mm-flux (7) inferred planet mass.}
	\label{tab:gaps}
	\begin{tabular}{lcccccc} 
		\hline
(1) Star name & (2) $\Delta/R$ & (3) $R$/au & (4) $M_\star/M_{\odot}$ & (5) $F_\nu$/mJy & (6) $M_{\rm dust}/M_{\rm Jup}$ &  (7) $M_{\rm p}/M_{\rm Jup}$\\
		\hline
RY Tau   & 0.129 &	43.41$\pm0.13$  &   $2.04^{+0.3}_{-0.26}$   &   210.39 &  0.29  & 0.077\\[3pt]
UZ Tau E & 0.115 &	69.05$\pm0.2$   &	$1.23^{+0.08}_{-0.08}$  &   129.52 &  0.19  & 0.023\\[3pt]
DS Tau   & 0.724 &	32.93$\pm0.32$  &	$0.83^{+0.02}_{-0.02}$  &   22.24  &  0.048 & 5.6\\[3pt]	
FT Tau   & 0.297 & 	24.78$\pm0.19$  &	$0.34^{+0.17}_{-0.09}$  &	89.77  &  0.12  & 0.15\\[3pt]
MWC480  & 0.329 & 73.43$\pm0.16$  &	$2.1^{+0.06}_{-0.06}$  &	267.76 &  0.59  & 1.3\\[3pt]
DN Tau   & 0.083 &	49.29$\pm0.44$  &	$0.87^{+0.17}_{-0.14}$  &	88.61  &  0.125 & 0.009\\[3pt]
GO Tau   & 0.239 &	58.91$\pm0.66$  &	$0.49^{+0.01}_{-0.01}$  &	54.76  &  0.097 & 0.057\\[3pt]
GO Tau   & 0.258 &	86.99$\pm0.88$  &	$0.49^{+0.01}_{-0.01}$  &	54.76  &  0.097 & 0.07\\[3pt]
IQ Tau   & 0.171 &	41.15$\pm0.63$  &	$0.74^{+0.01}_{-0.01}$  & 64.11  &  0.094 & 0.065\\[3pt]
DL Tau   & 0.182 &	39.29$\pm0.32$  &	$1.02^{+0.02}_{-0.02}$  &	170.72 &  0.37  & 0.11\\[3pt]
DL Tau   & 0.166 &	66.95$\pm0.87$  &	$1.02^{+0.02}_{-0.02}$  &	170.72 &  0.37  & 0.08\\[3pt]
DL Tau   & 0.262 &	88.9$\pm1.11$   &	$1.02^{+0.02}_{-0.02}$  &	170.72 &  0.37  & 0.33\\[3pt]
CI Tau   & 0.987 &	13.92$\pm0.32$  &	$0.91^{+0.02}_{-0.02}$  &	142.4  &  0.33  &  15.7\\[3pt]
CI Tau   & 0.281 &	48.36$\pm0.41$  &	$0.91^{+0.02}_{-0.02}$  &	142.4  &  0.33   & 0.36\\[3pt]
CI Tau   & 0.284 &	118.99$\pm0.65$ &	$0.91^{+0.02}_{-0.02}$  &   142.4  &  0.33   & 0.37\\[3pt]
		\hline
	\end{tabular}
\end{table*}

Numerical simulations of gas and dust are the best tool to constrain the planetary properties that reproduce a given structure in a disc. However, such numerical simulations are very time consuming to determine the planetary properties for our sizable sample of discs. Instead, we use empirically determined scaling relations between the gap properties and the planetary mass. In particular, for low viscosity discs ($\alpha\lesssim 0.01$), the gap width $\Delta$ (defined here as the distance between the location of the brightness minimum in the gap and the ring peak, see \citealt{Longetal18}) is expected to scale with the planet Hill radius 
\begin{equation}
R_{\rm H}=\left(\frac{M_{\rm p}}{3M_\star}\right)^{1/3}R, 
\label{eq:hill}
\end{equation}
where $R$ is the planet position (assumed here to coincide with the gap location), with a proportionality constant ranging from 4-8 depending on the disc parameters, so that $\Delta=kR_{\rm H}$ \citep{Dodson-robinson11,Pinillaetal12,Rosottietal16,fung16a,Facchinietal18}. 
Note that here we assume a one-to-one correspondence between a gap and a planet, while there is the possibility that multiple planets open a common single gap \citep{zhu11a} or that a single planet might open multiple gaps \citep{dong18c}. Finally, note that the gap width likely depends somewhat on disc hydrodynamical properties, such as pressure and viscosity \citep{Pinillaetal12,fung14a}.

Two discs in our sample, MWC480 \citep{Liuprep18} and CI Tau \citep{Clarkeetal18}, have been simulated with detailed hydrodynamical simulations to reproduce the gap properties.
MWC 480 presents a gap at $\sim 73$ au, which has been reproduced with a $2.3M_{\rm Jup}$ planet in the hydro simulations of \citet{Liuprep18}. The observed width of the gap in MWC 480 corresponds to $\sim 4.5 R_{\rm H}$. 
CI Tau presents three gaps at $\sim 14$, $48$ and $120$ au from the central star. Higher resolution observations of this system were obtained by \citet{Clarkeetal18}, who model the three gaps with three planets with $0.75$, $0.15$ and $0.4\,M_{\rm Jup}$. It should be noted that the gap widths observed in \citet{Clarkeetal18} are not easily comparable to the ones measured by \citet{Longetal18}, due to the different functional form of the radial dust profile used and in particular due to the fact that \citet{Clarkeetal18} use different inner and outer gap width, as opposite to the symmetrical Gaussian employed in \citet{Longetal18}. Despite these differences, the two outermost gaps appear to have a comparable normalized width in the two studies, while the innermost one is much larger in \citet{Longetal18} than in \citet{Clarkeetal18}. This discrepancy is probably due to the limited spatial resolution of our observations compared to \citet{Clarkeetal18} (at the distance of CI Tau, 19 au and 9 au, respectively) which is most important for the innermost ring, located at $\sim 14$ au. For consistency, in this paper we will always refer to the gap widths as measured by \citet{Longetal18}, keeping in mind that the width of the innermost gap in CI Tau might have been strongly overestimated. 

The width of the two outer gaps in CI Tau corresponds to $\sim 5$ and $7$ times the Hills radius of the planets used by \citet{Clarkeetal18} in their modeling. Thus, in the following, by averaging the results from hydrodynamical simulations of CI Tau and MWC 480, we will assume that the gap width $\Delta$ scales as 
\begin{equation}
    \Delta = 5.5R_{\rm H}.
    \label{eq1}
\end{equation}
We remind the reader that the relation above is related to the gap in the dust radial profile, that may be different than the gas gap (which we do not consider in this paper). 
The resulting planet masses calculated with Eq.~(\ref{eq1}) for the 15 gaps in our sample are reported in Table~\ref{tab:gaps}. 

The stellar masses are reproduced from those adopted by Long et al. (in preparation), obtained from a combination of dynamical mass measurements, when available \citep{simon00,pietu07,guilloteau14,simon17}, and otherwise by comparing literature estimates of temperature and luminosity to a combination of the \citet{baraffe15} and nonmagnetic models of \citet{feiden16}, as applied
by \citet{pascucci16}. UZ Tau E is a spectroscopic binary \citep[e.g.][]{prato02} and therefore has a dynamical mass that is much higher than would be expected from its spectral type.

In the plots shown below we also include error bars on the inferred planet masses coming from the uncertainty in the proportionality factor, ranging from 4.5 to 7 \citep{Rosottietal16}, resulting in an uncertainty in the inferred planet mass of the order of a factor $\sim 2$ either side, which dominates over the uncertainty on the assumed stellar mass.

Note that the outcome of hydrodynamical simulations of gas and dust with embedded planets depends on several physical and numerical parameters, including assumptions on the dust-gas coupling, the detailed treatment of the gas thermodynamics (locally isothermal equations of state are often used), the use of 2-dimensional or 3-dimensional codes, etc. All such assumptions imply an uncertainty in the relation between planet mass and width of the dust gap induced by it, often difficult to quantify. In this paper, we have simply assumed it to be given (see above) by the deviation between the different determination made by different groups using different codes and specific set-ups, although we warn that some of these uncertainties might be systematic (for example, most codes make the same assumptions on the thermodynamics, which may tend to overestimate the gap width for a given planet mass), and thus shared between all of the various simulations.

\section{Results}

\begin{figure}
	\includegraphics[width=\columnwidth]{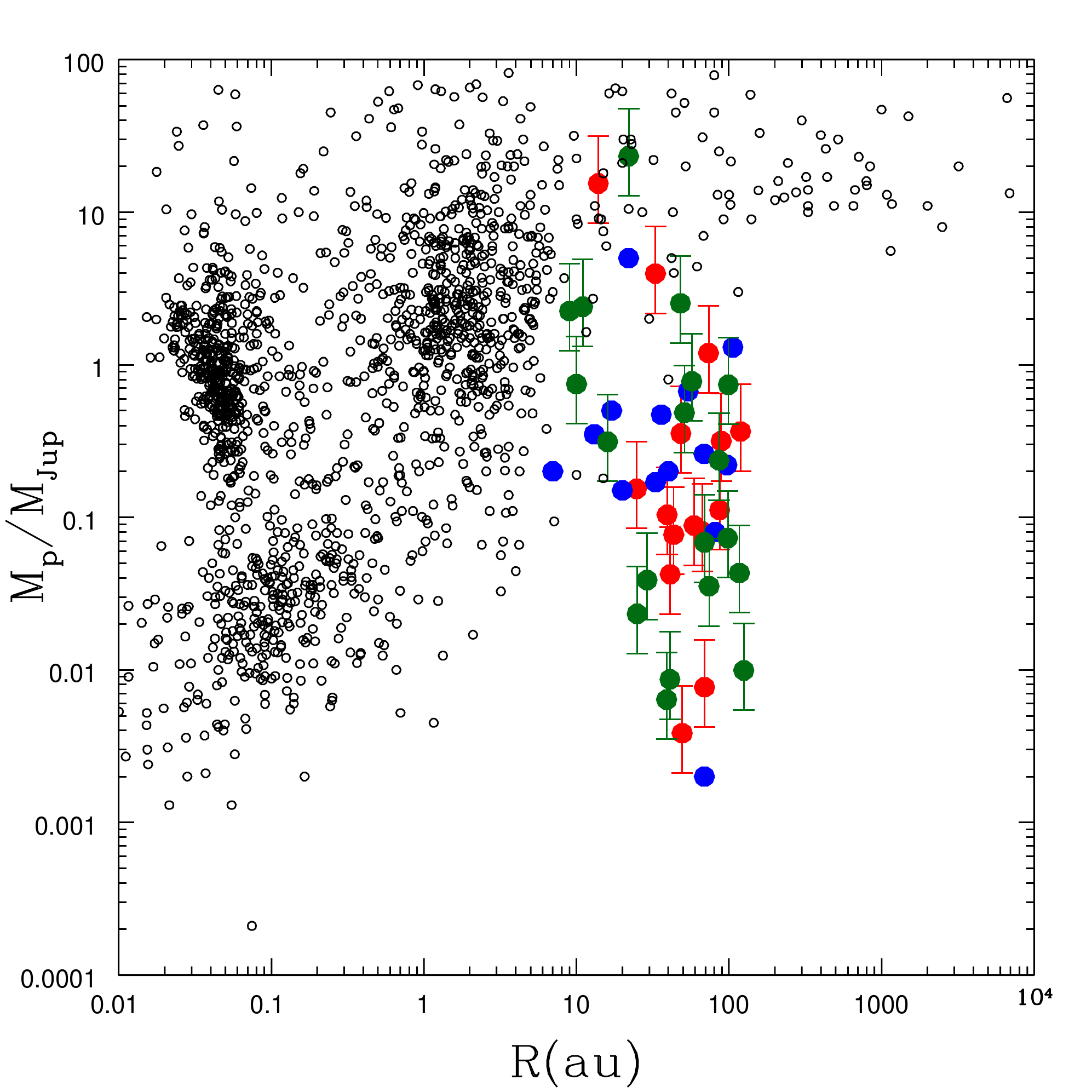}
    \caption{Plot of mass (y-axis) $M_{\rm p}$ vs. separation from the central star $R$ (x-axis) of the (empty circles) currently known exo-planets  (retrieved from the exoplanet.org database) compared to those obtained from the sample in \citet{Longetal18} (red points) and \citet{Zhang18} (green points) using Eq.~(\ref{eq1}), and those collected by \citet{Bae18} (blue points). Error bars in the planet masses indicate the uncertainty in the proportionality factor between gap width and planet's Hills radius, assumed to be in the range $[4.5-7]$.}
    \label{fig:mass_vs_a}
\end{figure}

Figure~\ref{fig:mass_vs_a} shows a comparison between masses and locations of currently known exo-planets (empty circles, data from www.exoplanet.eu, as of the 31st of Octobr 2018) and those inferred from the gap extents in \citet{Longetal18} (red points) using Eq.~(\ref{eq1}). 
Recently, the DSHARP ALMA Large Program data have been released, with an analysis of additional gaps in bright protostellar discs. \citet{Zhang18} measured the width\footnote{Note that \citet{Zhang18} define the gap width in a slightly different way than us, so that $\Delta_{\rm Zhang}/R=(R_{\rm out}-R_{\rm in})/R_{\rm out}$, where $R_{\rm out,in}$ are the outer/inner radius of the gap, which makes their gap size of the order of two times the one obtained with our definition. When using their sample, we have corrected their data for this difference.} of 19 gaps, from which we calculate the putative planet mass with the same procedure as we used for the \cite{Longetal18} sample, with stellar and disc parameters taken from \citet{Zhang18}. The resulting planet masses are shown with green points in Fig.~\ref{fig:mass_vs_a} and are listed in Table \ref{tab:dsharp}. Despite the differences in estimating the planet masses, they appear to be consistent with those quoted by \citet{Zhang18}. 

In addition, we also plot as blue circles the planet masses and locations inferred from other 14 ringed discs and disc hosting cavities (so called transition discs), as collected by \citet{Bae18} (see their Fig. 1). For the few cases (HD163296, Elias 24 and AS209) that are present both in the DSHARP and in the \citet{Bae18} sample, we use the planet mass obtained from the measured gap width in DSHARP. We list the location and mass of the planets collected by \citet{Bae18} in Table \ref{tab:bae}. In total, we thus have 48 planets inferred from the gaps in dusty discs, that is the largest gap sample analysed to date.

\begin{table}
	\centering
	\caption{Planet masses for the gaps in the DSHARP survey \citep{Zhang18}. The columns indicate, respectively: (1) star name; (2) Gap width according to \citet{Zhang18} ; (3) Gap location; (4) inferred planet mass.}
	\label{tab:gaps_dsharp}
	\begin{tabular}{lccc} 
		\hline
		(1) Star name & (2) $\Delta_{\rm Zhang}/R$ & (3) $R$/au & (4)  $M_{\rm p}/M_{\rm Jup}$\\
		\hline
AS209       & 0.42 & 9 &	2.25\\
AS209       & 0.31 & 99& 	0.74\\
Elias 24    & 0.32 & 57& 	0.77\\	
Elias 27    & 0.18 & 69& 0.07\\
GW Lup      & 0.15 & 74& 	0.035\\
HD 142666   & 0.2  & 16& 	0.3\\
HD 143006   & 0.62 & 22& 	23\\
HD 143006   & 0.22 & 51& 	0.48\\
HD 163296   & 0.24 & 10& 	0.74\\
HD 163296   & 0.34 & 48& 	2.5\\
HD 163296   & 0.17 & 86& 	0.23\\
SR4         & 0.45 & 11& 	2.4\\
DoAr 25     & 0.15 & 98&	0.07\\
DoAr 25     & 0.08 & 125&	0.01\\
Elias 20    & 0.13 & 25 &  0.02\\
IM Lup      & 0.13 & 117&  0.04\\
RU Lup      & 0.14 & 29 &	0.038\\
Sz 114      & 0.12 & 39 &	0.006\\
Sz 129      & 0.08 & 41 &	0.008\\
		\hline
	\end{tabular}
	\label{tab:dsharp}
\end{table}

\begin{table}
	\centering
	\caption{Planet masses collected by \citet{Bae18}. The columns indicate, respectively: (1) star name; (2) Gap location; (3) inferred planet mass.}
	\label{tab:gaps_bae}
	\begin{tabular}{lcc} 
		\hline
		(1) Star name & (2) $R$/au & (3) $M_{\rm p}/M_{\rm Jup}$\\
		\hline
HL Tau      & 13.1 & 0.35\\
HL Tau      & 33   & 0.17\\
HL Tau      & 68.6 & 0.26\\	
TW Hya      & 20   & 0.15\\
TW Hya      & 81   & 0.08\\
HD 169142   & 54   & 0.67\\
HD 97048    & 106  & 1.3\\
Lk Ca 15    & 36   & 0.47\\
RXJ 1615    & 97   & 0.22\\
GY 91       & 7    & 0.2\\
GY 91       & 40   & 0.2\\
GY 91       & 69   & 0.002\\
V 4046      & 17   & 0.5\\
PDS 70      & 22   & 5\\
		\hline
	\end{tabular}
	\label{tab:bae}
\end{table}

The inferred planet masses from our sample and the \citet{Zhang18} sample are consistent with those of the \citet{Bae18} sample, although we caution that the method used to derive them are significantly different: while the masses collected by \citet{Bae18} are mostly inferred from hydrodynamical simulations, coupled with a dust evolution module, our estimates are based on a simpler approach. It is interesting to note, however, that the two approaches lead to compatible results.

The properties of the putative planets obtained with our method populate a region in the mass vs. separation diagram that cannot be probed by the current exo-planet surveys. 
We note that the observations of planets at distances $\gtrsim 10\, {\rm au}$ from the central star are biased toward large masses: at those separations planets can be detected mostly by direct imaging or by microlensing. Recent determinations of the occurrence rates of massive planets ($M>2M_{\rm Jup}$) beyond 10-20 au are in the range of a few up to 5\% \citep{Bowler18}. More specifically, the 68\% confidence interval is estimated to be $[1.6-5.1]$\% for $2-14M_{\rm Jup}$ planets between 8 and 400 au by \citet{Lannier16}, $[4-10]$\% for $5-20M_{\rm Jup}$ planets between 10 and 1000 au by \citet{Meshkat17} and $[0.75-5.7]$\% for $0.5-75M_{\rm Jup}$ between 20 and 300 au by \citet{Vigan17}. Note, however, that such estimates suffer from very large uncertainties, depending on whether one uses a hot or a cold start model for the planet. For example, \citet{Stone18}, using a cold start model, put an upper limit to the occurrence rate of $7-10M_{\rm Jup}$ planets between 5 and 50 au as high as 90\% for FGK stars.

For the combined sample, including the \citet{Longetal18}, the \citet{Zhang18} and the  \citet{Bae18} data the occurrence rate of such massive planets is $7/48\sim 15$\%, which is slightly higher than the published rates. However, note that, apart from the \citet{Longetal18} sample, the other gap detections all present strong biases to very luminous mm sources. Furthermore, it is important to note that these planets will naturally accrete mass and migrate to the inner disc during their evolution, and thus change their properties, see Sect. 3.1.

\begin{figure}
	\includegraphics[width=\columnwidth]{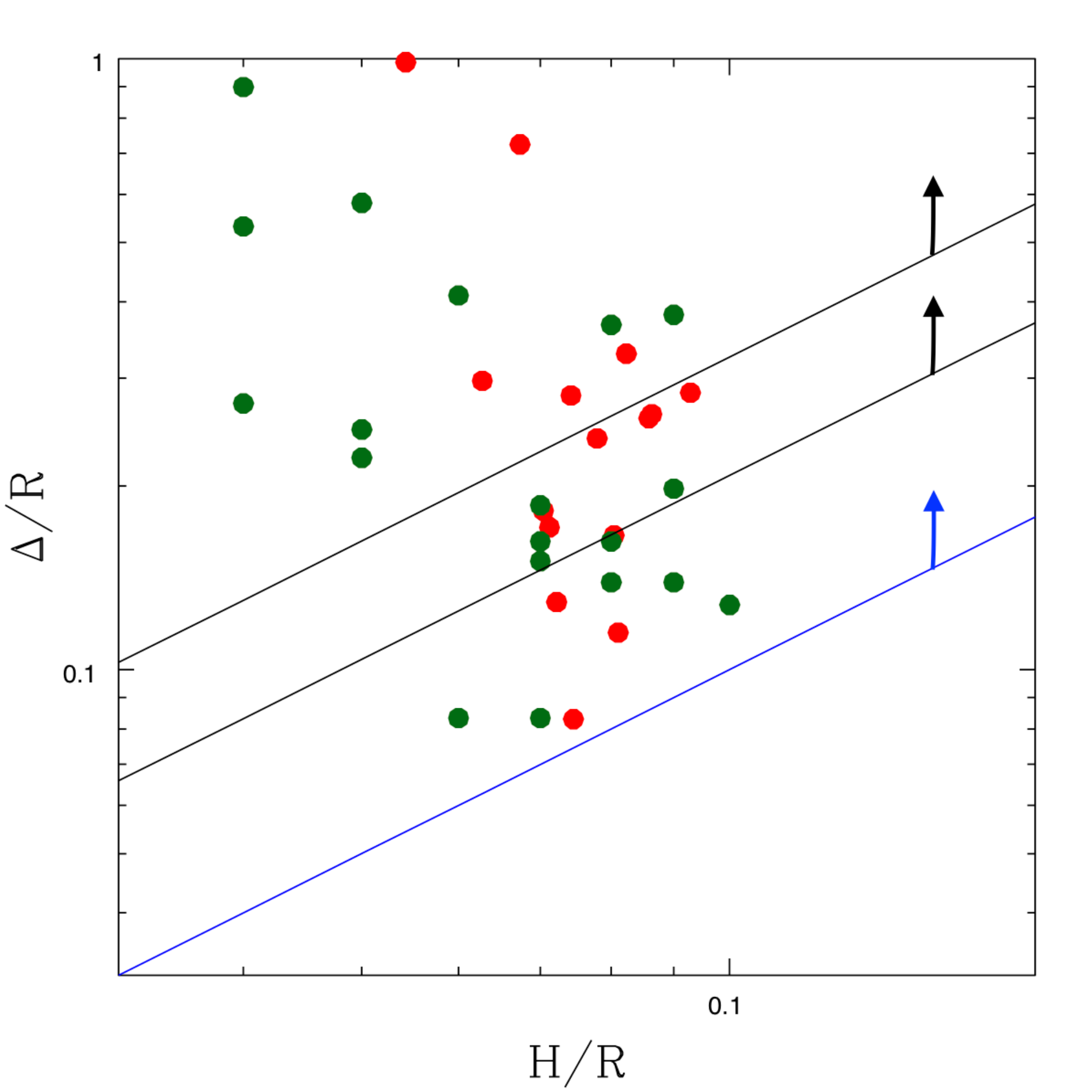}
    \caption{Measured gap widths vs. disc aspect ratio (as estimated from Eq. (\ref{eq:honr})) in the \citet{Longetal18} (red points) and in the \citet{Zhang18} (green points) samples. The two black lines indicate the range $[2.1-3.2]H/R$ above which the gap width is expected to lie if the dust is strongly coupled to the gas ($\mathrm{St}\ll 1$). The blue line indicates the relation $\Delta=H$, that is the minimum gap width expected for dust with $\mathrm{St}\sim 1$.}
    \label{fig:delta_vs_h}
\end{figure}

From the planet-disc interaction point of view, the minimum planet-star mass ratio able to carve a dust gap depends on the coupling between the gas and the dust, as measured by the Stokes number 
\begin{equation}
\mathrm{St}=\Omega\, t_{\rm stop},
\end{equation}
where $t_{\rm stop}$ is the drag stopping time and $\Omega$ is the local Keplerian frequency \citep{Weidenschilling77}. In particular, for strongly coupled dust grains (with $\mathrm{St}\ll 1$) the minimum dust gap opening planet mass is  
\begin{equation}
\frac{M_{\rm min}}{M_\star}=0.3\left(\frac{H}{R}\right)^3,
\label{eq:condigapdust}
\end{equation}
where $H/R$ is the disc aspect ratio at the planet position, which depends on the disc temperature \citep{lambrechts14a,Rosottietal16,DipierroLaibe2017}. If we consider a standard irradiated disc model \citep{ChiangGoldreich1997,Dullemond02,Armitage2010}, the disc aspect ratio is given by
\begin{equation}
\frac{H}{R}\approx 0.05\left(\frac{R}{10\mbox{au}}\right)^{1/4}.
\label{eq:honr}
\end{equation}
In practice, since we obtain the planet mass from the gap width by assuming that it scales with the planet Hill's radius, the condition $M_{\rm p}\gtrsim M_{\rm min}$ implies (through Eqs.~\ref{eq:hill} -\ref{eq:condigapdust}) that 
\begin{equation}
\frac{\Delta}{R}\gtrsim [2.1-3.2]\,\frac{H}{R}
\end{equation}
for strongly coupled dust, where the brackets correspond to our chosen interval in the proportionality factor in Eq.~(\ref{eq:hill}) ($k=[4.5-7]$). 
For more loosely coupled dust grains ($\mathrm{St}\gtrsim 1$), conversely, a dust gap can be opened relatively more easily because viscous and pressure forces are not effective in closing the gap. Combining Eqs. (56) and (58) in \citet{DipierroLaibe2017}, we obtain in this case the requirement:
\begin{equation}
\frac{\Delta}{R}\gtrsim \mathrm{St}^{-1/2}\frac{H}{R}.
\end{equation}
Note that, for $\mathrm{St}\leq 1$ the gap width cannot be smaller than the disc thickness $H$. 

In Fig.~\ref{fig:delta_vs_h} we plot the gap width $\Delta/R$ for the gaps in the two samples of \citet{Longetal18} (red points) and \citet{Zhang18} (green points) versus the disc aspect ratio at the gap location $H/R$, as computed from Eq.~(\ref{eq:honr}). The two black lines indicate the range $[2.1-3.2]H/R$ above which we should expect the gap width to lie, if the dust is strongly coupled to the gas. The blue line shows instead the simple relation $\Delta=H$, that is the minimum gap width expected for dust with $\mathrm{St}\sim 1$. As we can see, most of our points are consistent with the dust being strongly coupled to the gas. In a few cases the gap width appears to be somewhat smaller, which may imply that in these systems the dust is less coupled and it is thus easier to open up a dust gap. 

\begin{figure}
	\includegraphics[width=\columnwidth]{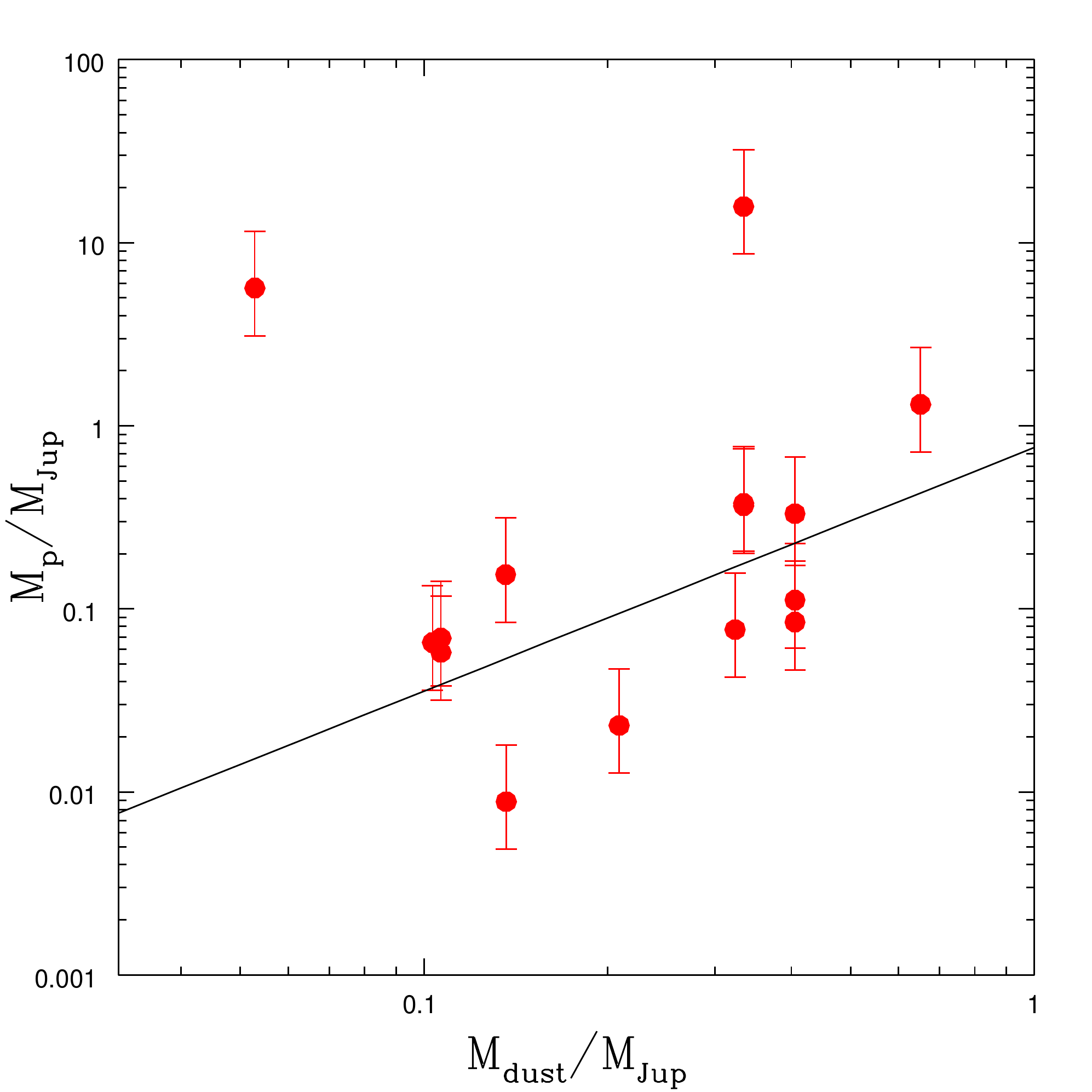}
    \caption{Mass of the planets $M_{\rm p}$ (y-axis) versus total dust mass in the disc (x-axis) for the putative planets in \citet{Longetal18}. The solid line indicates the linear regression of the form $M_{\rm p}\propto M_{\rm dust}^{1.33}$.}
    \label{fig:mass_vs_flux}
\end{figure}

\begin{figure}
	\includegraphics[width=\columnwidth]{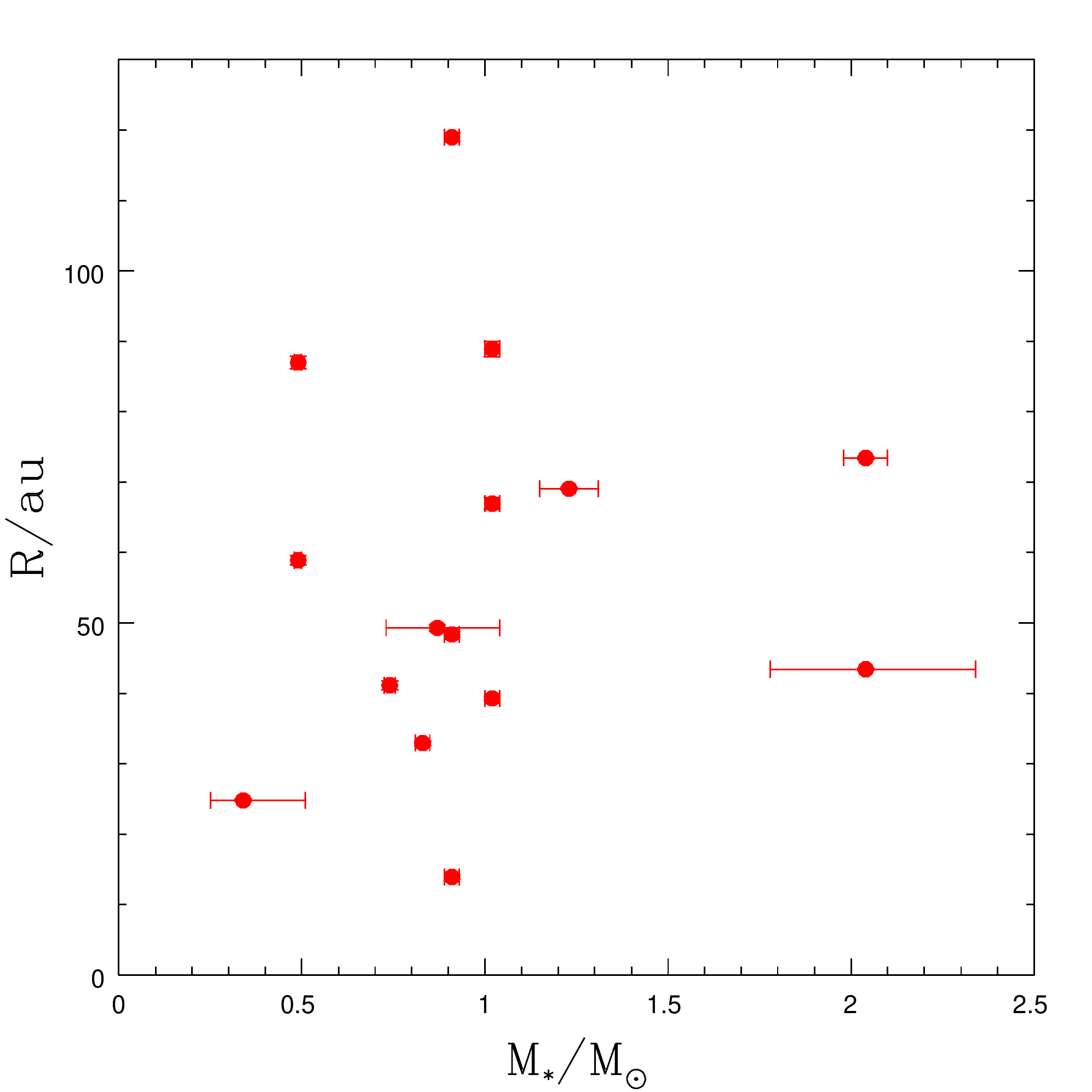}
    \caption{Scatter plot showing the gap location $R$ (y-axis) vs. the central star mass $M_\star$ (x-axis) for the putative planets in \citet{Longetal18}.}
    \label{fig:a_vs_starm}
\end{figure}

Next, we check for possible correlations between the derived planet mass and the disc dust mass, as measured from the mm flux, assuming optically thin emission, a dust temperature of $T_{\rm dust}=20$K and a dust opacity\footnote{Although note that the dust opacity values are very uncertain, as it depends on the local size distribution and composition of dust grains, that is controlled by grain growth and radial drift.} $\kappa=2.3 \,(\nu/230\mbox{GHz})^{0.4}$ cm$^2/$g. This is plotted in Fig.~\ref{fig:mass_vs_flux}, which shows the mass of the putative planets versus the total dust mass in the disc \citep{Longetal18}. Apart from the two most massive planets (corresponding to the inner ring of CI Tau and to DS Tau), the rest of our small sample appears to follow a tentative trend. The solid line in Fig.~\ref{fig:mass_vs_flux} shows the best linear regression of the data (excluding the two outliers) in the form $M_{\rm p}\propto M_{\rm dust}^{1.33}$. Note that, of course, this plot relates the planet mass to the \emph{current} dust mass in the disc, which does not necessarily represent a proxy for the disc mass \emph{at the time of planet formation} \citep{nixon18a}. 
Moreover, inferring the value of the dust mass from continuum observations of protoplanetary discs is still under debate, mostly due to uncertainty in dust opacity and optical depth \citep{bergin18a}.
Indeed, \citet{Manara18}, using photometric data, have recently shown that the disc dust masses measured from mm fluxes may be in general lower than the mass of exo-planets  (but see \citealt{Mulders15} and \citealt{pascucci16} for a different opinion, based on Kepler planet mass estimates), as also confirmed by spatially resolved studies \citep{Tazzari17}, who find dust surface density profiles below the Minimum Mass Solar Nebula in their Lupus disc sample. This can be explained with either a rapid formation of planetary cores \citep{Najita14}, or a replenishment of the disc from the environment, or a sizable fraction of circumstellar dust being captured in larger dust agglomerations such as boulders, planetesimals, etc. Especially for the two most massive inferred planets in our sample, it is possible that most of the primordial disc mass might have already ended up in planets, that thus might appear to live in less massive discs than the correlation would suggest.

In a sample of transition discs, \citet{Pinillaetal18} did not find any correlation between mm-flux and cavity size. Note that although also in transition discs the cavity is sometimes interpreted as the effect of the presence of a planet, here we are not concerned with discs with cavities, but only in gaps. 

Finally, in Fig.~\ref{fig:a_vs_starm} we show the location of the gaps in our sample versus the stellar masses. No clear trend can be recognised here, indicating that, in the planet interpretation, the planet formation region does not appear to depend strongly on the stellar mass.

\subsection{The fate of planets}
\label{sect:fatepla} 
Due to interactions between planets and the surrounding disc material, the properties of the putative planets inferred in gapped-like discs around young stellar objects are expected to evolve with time. As a result, the planets would generally migrate and accrete mass from the surrounding disc. 

In order to predict if the planets will survive to their migration and to compare their final properties with those of currently known exo-planets, we compute the variation of the separation and mass of the planets under consideration using prescribed migration and accretion laws, assuming that the disc properties are fixed in time.
We assume that the planets migrate according to type I or type II migration regime (e.g., see \citealt{Papaloizou06}), depending on their ability to carve a deep gap in the local gas density structure (as opposed to the dust gaps that we know have been opened in all of our putative planets). Starting from the initial properties of the planets (see Tables~\ref{tab:gaps}, \ref{tab:gaps_dsharp} and \ref{tab:gaps_bae}), we assume that the gap-opening mass $M_{\mathrm{p,gap}}$ in the gas disc is given by the \citet{crida06a} criterion, corresponding to a drop of the local gas surface density to a factor $\sim 10\%$ of the unperturbed value, i.e.
\begin{equation}
\frac{3}{4}\frac{H}{R_{\mathrm{H}}}+\frac{50\,\alpha M_{\star}}{M_{\mathrm{p,gap}}} \left(\frac{H}{R}\right)^2= 1 ,
\label{eq:condcrida}
\end{equation}
where  $\alpha$ indicates the Shakura-Sunyaev turbulence parameter \citep{Shakurasunyaev1973}, assumed to be equal to $0.005$ \citep{flaherty17a}. The value of the aspect ratio at the planet position is obtained from Eq.~(\ref{eq:honr}). We adopt a simplistic bimodal model for planetary migration by assuming that planets with mass smaller (larger) than $M_{\mathrm{p,gap}}$ migrate according to type I (II) regime. 

\begin{figure}
	\includegraphics[trim={0.2cm 0cm 0cm 0.1cm},clip,width=\columnwidth]{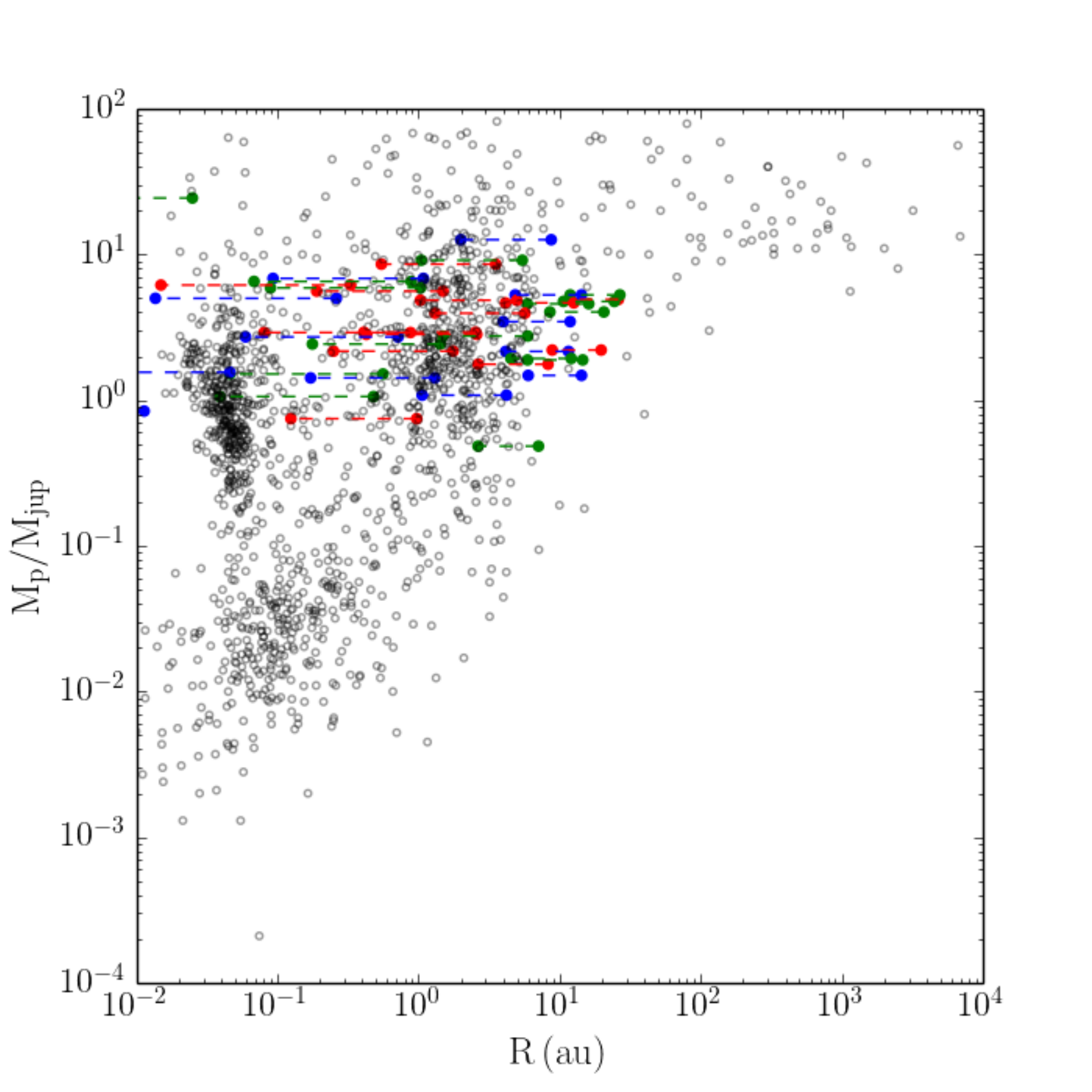}
    \caption{Same as Fig.~\ref{fig:mass_vs_a} but where the points indicate the expected final mass and location of the putative planets inferred in \citet{Longetal18} (red), \citet{Zhang18} (green) and \citet{Bae18} (blue) after 3 and 5 Myr of planet evolution. The dashed lines indicate the range of planet locations after a total time in the range [3,5] Myrs. Planetary accretion and migration lead to a redistribution of planet properties that mostly populates the branch of cold Jupiters.}
    \label{finalp_finalm}
\end{figure}

The planet orbital evolution and accretion history is then computed following the method of \citet{Dipierroetal18} (see their Sec. 4.4 for details). In particular, we assume that low mass planets (i.e. $M_{\mathrm{p}} < M_{\mathrm{p,gap}}$) initially undergo a rapid growth and migration phase (corresponding to the Type I regime, when the planet is still embedded in the disc), rapidly reaching a mass and radius given by Eqs.~(20)-(22) in \citet{Dipierroetal18}.
Then, we let the planets migrate without growing in mass on the slower viscous timescale of the disc:
\begin{equation}
t_{\mathrm{migr,II}}=\frac{2}{3}\left(\frac{1}{\alpha \Omega} \right) \left(\frac{H}{R} \right)^{-2}.
\end{equation}
Those planets in our sample with an initially high mass (i.e. $M_{\mathrm{p}} > M_{\mathrm{p,gap}}$) simply migrate toward the central star according to the type II regime. If the planet mass is much larger than the local disc mass, Type II migration is expected to be further slowed down by a factor $B=M_{\rm p}/4\pi\Sigma R^2$, where $\Sigma$ is the total (gas+dust) disc surface density \citep{ivanov99a}. However, given that the dust masses for our sample (see Table 1) are generally of the order of the estimated planet mass, and assuming a gas-to-dust ratio of 100, we find that none of our planets is massive enough to be in this modified Type II migration regime.

Fig.~\ref{finalp_finalm} shows the final properties (separation from the central star and mass) of the planets in our sample.
The dashed lines indicate the range of planet locations after a total time in the range [3,5] Myrs (taken to be an estimate of the gas disc lifetime, including a possible spread in ages and evolutionary time), compared to those inferred from the currently known exo-planets. Initially, around half of the planets in our sample have a mass below the one given by the gas gap-opening criterion and therefore accrete mass and migrate in type I regime. We find that these migrating and accreting planets will reach the gap opening mass (Eq.~\ref{eq:condcrida}) and transit into the slow type II migration regime well before being lost into the central star (and thus save themselves from rapid migration), consistently with recent findings of \citet{crida17a} and \citet{johansen18a}. More massive planets (i.e. $M_{\mathrm{p}} > M_{\mathrm{p,gap}}$) simply slowly migrate toward the central star according to the type II regime. After planetary migration and accretion, $\sim 20$\% of the planets are lost into the star (we assume that a planet is lost into the star if its separation is smaller than 0.01 au). Moreover, nearly all of the planets in our sample reach a mass above Jupiter.

Our evolutionary model is very simplified and approximated: we have kept the disc properties fixed during the evolution, we have simply assumed a uniform lifetime for all the discs (neglecting also a possible range in ages in our sample) and we have neglected possible modifications to the migration laws \citep[e.g.][]{ivanov99a,Durmann15}. However, it is interesting to note that the final distribution of the planets is consistent with the known properties of the exo-planet population, especially those placed in the branch of cold Jupiters. 

Since the planetary growth and migration are closely linked to the disc evolution, a proper investigation should take into account the underlying evolution of the dynamical and thermal structure of the gas and dust content in protoplanetary discs, along with the possible presence of mechanisms acting to slow-down (or even reverse) the inward planet migration such as photoevaporation \citep{Matsuyama03,alexander12a}, migration in a multiple planet system \citep{Martin07}, disc migration feedback \citep{fung18a}, sublimation lines, shadowed regions and heat transition barriers \citep[e.g.][]{bitsch15a,baillie16a,ndugu18a,johansen18a}, and even further migration occurring by planet-planet interaction after the disc is dispersed. 

\section{Conclusions}

In this paper we have analysed the sample of rings and gaps observed to date in protoplanetary discs to infer the properties of the population of planets that might have been able to carve the observed gaps. 
Our analysis includes the recent detections of gaps in discs in the Taurus star forming region by \citet{Longetal18}, along with the recent observations in the DSHARP ALMA Large Program analysed by \citet{Zhang18} and the additional sample of gaps collected by \citet{Bae18}.
For those discs where a proper hydrodynamical modelling was not carried out to infer the planet properties, we estimate the putative planet masses assuming that the gap width is proportional to the planet Hill's radius. We then describe some possible correlations of the putative planet properties with the other system parameters.

The most important conclusion of our work is that there appears to be no discrepancy between the possibility that embedded planets are responsible for carving gaps in discs around young stars and the lack of detections at similar locations by dedicated planet searches. First, we find that the locations and masses of the planets around these young stars occupy a distinct region in the planet mass versus semi-major axis plane that is presently not probed by planet detection campaigns (around both young, T Tauri stars and older, main sequence stars). The high frequency of gaps observed in planet forming discs has sometimes been interpreted as evidence against a planet induced model for gap formation, based on the fact that planet detection campaigns do not observe massive planets at tens of au very frequently. Our analysis, however, shows that if the planets remain at the lower end of the masses required to create gaps then they would be, as yet, undetectable by campaigns searching at these distances.  

The number of gaps in the sample of \citet{Longetal18} (which is the least biased sample of gaps in discs available so far) is 15 out of 32 targets. Taking into account the fraction of disc hosting stars in Taurus, which is 0.75 \citep{Luhman10}, this leads to an occurrence rate of gaps around young stars of 35\%. \citet{fernandes18a} have compared favourably this number with their estimate of the number of giant planets (with masses in the $[0.1-20]M_{\rm Jup}$ range and semi-major axis in the $[0.1-100]$ au range), which is 26.6\%. A similar occurrence rate from RV surveys has also been published by \citet{cumming08a}, who estimate a value of 17-20\% for giant planets (above Saturn mass) within 20 au. However, one must remember that the occurrence rates of giant planets from RV surveys or direct imaging should not be directly compared with the occurrence rates of gaps, because planets migrate and accrete mass during the disc evolution.

Motivated by this, we further explore the final properties of the planets in our sample by using a simple prescription of planetary migration and accretion \citep{Dipierroetal18}. After 3-5 Myr of planetary evolution, we find that the final properties of the planets approach the branch of cold Jupiters in the current observed distribution of exoplanets. Thus, planetary migration and accretion provides a second explanation for the lack of detected planets at large distances around older, main sequence stars. 

After planetary migration and accretion, $\sim 20$\% of the planets are lost into the star. However, for the sub-sample including only the \citet{Longetal18} discs, only one planet is lost and the final number of surviving planets is 14, most of them having masses above Jupiter. In total, thus, the occurrence rate of Jupiter mass planets in our model is 33\%. As mentioned above, \citet{fernandes18a} estimate a value of 26.6\% for the occurrence rate of giants (with masses above $0.1M_{\rm Jup}$), but this number is reduced to only 6\% for Jupiter mass planets, according to \citet{fernandes18a}. This interesting fact can be explained in several different ways. First, we note that our estimates are certainly affected with low-number statistic uncertainties, and future, unbiased larger surveys should improve in this respect. Second, it is worth noting that planet detection campaigns concentrate on Solar type stars, while this is not the case for the disc surveys, which include a wider range of stellar types. Third, our planetary accretion model probably overestimates the amount of accreted mass. Indeed, we assume an isothermal equation of state to compute the accretion rate \citep{Dipierroetal18}, which is the maximum accretion rate allowed \citep{ayliffe09a,Szulagyi15,szulagyi16a,lambrechts17a}.  Certainly, this kind of comparison can put interesting constraints on accretion and migration models.

Estimating the presence of a planet based on the gap it carves in the protoplanetary disc naturally has a bias in that very low mass planets do not induce gaps. Such a bias can be quantified using known relationships between the minimum gap opening planet mass (and thus the minimum expected gap width) and the disc aspect ratio. Our results show that the measured gap widths are generally larger than a few times the disc thickness $H$, which is consistent with predictions for planet gap opening for a dust population strongly coupled to the gas. In a few cases, the gap width is comparable to $H$, which might imply that the dust-gas coupling in these systems is lower. However, these gaps are still consistent with $St\gtrsim 1$ for the mm-sized grains, as required for them to remain at their current location and not undergo rapid inward drift. In no cases do we find gap widths smaller than H, which strongly supports our hypothesis that the observed gaps are opened by planets.

Despite uncertainties coming from the small size of our sample (and by the presence of a couple of outliers), we suggest that there could be a correlation between planet mass and disc mass, as inferred from the disc mm flux, supporting the notion that more massive discs tend to produce more massive planets. However, note that a similar correlation between mm flux and cavity size was not found for the larger cavities (as opposed to the gaps discussed here) around transition discs, analysed by \citet{Pinillaetal18}.
No correlation is instead found between the location of the gaps and the stellar mass, possibly indicating that the planet formation region does not appear to depend strongly on stellar mass, although again note that this might be affected by the relatively small sample size. 

Upcoming surveys of discs will certainly add more data points to our currently small sample and further refine or reject our findings. Importantly, theoretical models have developed to the point of making a priori predictions for exo-planet demographics.
In general, analyses such as ours, once the samples are more complete, will be needed to relate the properties of newborn planets with the ``adult'' planet population coming from planet detection campaigns around main sequence stars, thus posing important constraints on the early evolution of planets in their discs.

\section*{Acknowledgements}

We thank an anonymous referee for their insightful comments. We thank Jaehan Bae for providing us with masses and separations from the central star collected in \citet{Bae18}.
GL, ERa, BN and ERi acknowledge support by the project PRIN-INAF 2016  
The Cradle of Life - GENESIS-SKA (General Conditions in Early 
Planetary Systems for the rise of life with SKA). This project has received funding from the European Union's Horizon 2020 research and innovation programme under the Marie Sklodowska-Curie grant agreement No 823823 (Dustbusters RISE project). GD and ERa acknowledge financial support from the European Research Council (ERC) under the European Union's Horizon 2020 research and innovation programme (grant agreement No 681601). 
MT has been supported by the DISCSIM project, grant agreement 341137 funded by the European Research Council under ERC-2013-ADG. DJ is supported by the National Research Council of Canada and by an NSERC Discovery Grant. FMe, GvdP, YB acknowledge funding from ANR of France (contract ANR-16-CE31-0013, Planet-Forming-Disks). FL and GJH are supported by general grants 11773002 and 11473005 awarded by the National Science Foundation of China. CFM acknowledges an ESO Fellowship and was partly supported by the Deutsche Forschungs-Gemeinschaft (DFG, German Research Foundation) - Ref no. FOR 2634/1 TE 1024/1-1, by the DFG cluster of excellence Origin and Structure of the Universe (www.universe-cluster.de)




\bibliographystyle{mnras}

\bibliographystyle{mnras}
\bibliography{biblio}







\bsp	
\label{lastpage}
\end{document}